\documentclass[10pt]{article}
\usepackage[latin1]{inputenc}
\usepackage{amsmath}
\usepackage{amsfonts}
\usepackage{amssymb}
\usepackage{graphicx}
\usepackage{epstopdf}
\usepackage{color}
\usepackage{appendix}
\usepackage{csvsimple}
\usepackage{longtable}
\usepackage{chngpage}
\usepackage{subcaption}
\usepackage[export]{adjustbox}[2011/08/13]
\usepackage{mathrsfs}
\usepackage{a4wide}
\usepackage{psfrag}
\usepackage{multirow}
\usepackage{hhline}
\usepackage{grffile}
\begin{document}
	
\title{\bf\boldmath Quintessential Inflation with $\alpha$-attractors}

\author{Konstantinos Dimopoulos and Charlotte Owen\\
	\\
	{\small\em Consortium for Fundamental Physics, Physics Department,}\\
	{\small\em Lancaster University, Lancaster LA1 4YB, UK}
	\\
	\vspace{-.4cm}
	\\
	{\small e-mails: {\tt k.dimopoulos1@lancaster.ac.uk}, \
		{\tt c.owen@lancaster.ac.uk}}}
\vspace{1cm}

\maketitle

\begin{abstract}
A novel approach to quintessential inflation model building is studied, 
within the framework of $\alpha$-attractors, motivated by supergravity theories.
Inflationary observables are in excellent agreement with the latest CMB 
observations, while quintessence explains the dark energy observations without
any fine-tuning. The model is kept intentionally minimal, avoiding the 
introduction of many degrees of freedom, couplings and mass scales. 
In stark contrast to $\Lambda$CDM, for natural values of the parameters,
the model attains transient accelerated expansion, which avoids the future 
horizon problem, while it maintains the field displacement mildly sub-Planckian 
such that the flatness of the quintessential tail is not lifted by radiative 
corrections and violations of the equivalence principle (fifth force) are
under control. In particular, the required value of the cosmological constant 
is near the eletroweak scale. Attention is paid to the reheating of the 
Universe, which avoids gravitino overproduction and respects nucleosynthesis 
constraints. Kination is treated in a model independent way. A spike in 
gravitational waves, due to kination, is found not to disturb nucleosynthesis 
as well.
%is argued to be potentially observable by LIGO in the near future.
\end{abstract}

\nopagebreak

\section{Introduction}\leavevmode

The Universe is currently in a phase of accelerated expansion. Since the 
observational discovery of this from type Ia Supernovae \cite{SN}, it has been 
confirmed by several methods, most notably observations of the Cosmic 
Microwave Background (CMB) \cite{WMAP,planck,BICEP2}. The observed dynamics can
only be explained via the introduction of some hypothetical substance, called
dark energy (for a comprehensive review see Ref.~\cite{review}). The simplest
form of dark energy, which does not require new Physics, is a non-zero 
cosmological constant corresponding to positive vacuum density. However,
since this vacuum density must be comparable to the present density of the 
Universe (accelerated expansion started in the last billion years only) 
the value of the vacuum density must be incredibly fine-tuned, down to
$\sim 10^{-120}M_P^4$, where $M_P = 1.22 \times 10^{19}\mathrm{GeV}$ is the 
Planck mass, which is the natural scale of gravity in General Relativity 
\cite{weinberg}. Unless introducing a new scale in Physics and a new hierarchy 
problem, this explanation of the observed recent accelerated expansion seems 
unnatural, especially since the dark energy component is thought to make up 
almost 70\% of the current content of the Universe. 

Following the success of the inflationary paradigm,
a promising alternative to explain the late time acceleration of the Universe 
is a dynamic scalar field, provided that it avoids the extreme fine-tuning of 
the cosmological constant. This field has been referred to as quintessence; the 
fifth element of the current make up of the Universe, after baryonic and cold 
dark matter, radiation and neutrinos \cite{Q}. Quintessence can generate the
observed accelerated expansion if it dominates the Universe at present, while
rolling down a flat potential, in the same way as the inflaton field drives 
inflation in the early Universe. However, quintessence suffers from its own 
tuning problems. Indeed, in fairly general grounds it can be shown that 
quintessence needs to travel at least over Planckian distances in field space 
whilst retaining the flatness of its potential against radiative corrections. 
Also, the effective mass of quintessence is comparable to the Hubble constant 
\mbox{$H_0=1.43\times 10^{-33}\,$eV}, such that its Compton wavelength is 
comparable to the present horizon. Consequently, if not suppressed, interactions
of the quintessence field with the standard model can correspond to a 
long-range `fifth force', which may result in violations of the equivalence 
principle \cite{5th}. In addition, the introduction of yet another 
unobserved scalar field (on top of the inflaton field) seems unappealing. 
Finally, a rolling scalar field introduces another tuning problem, namely that 
of its initial conditions.

A compelling way to overcome the difficulties of the quintessence scenario is 
to link it with the rather successful inflationary paradigm. This is quite 
natural since both inflation and quintessence are based on the same idea; that 
the Universe undergoes accelerated expansion when dominated by the potential 
density of a scalar field, which rolls down its almost flat potential. This 
unified approach has been named quintessential inflation \cite{qinf}.
In quintessential inflation the scalar potential is such that it causes two 
phases of accelerated expansion, one at early and the other at late times.
Apart from using a single theoretical framework to describe both inflation and 
dark energy, quintessential inflation overcomes the problem of initial 
conditions of quintessence, because they are determined by the inflationary
attractor.

Modelling quintessential inflation is not easy. The two plateaus featured in 
the potential are bridged by a steep dip over more than a hundred orders of 
magnitude. Yet, there have been many early attempts \cite{kination,QIearly,eta}
and since then, the subject has continued to be investigated \cite{QIlate}. 
In quintessential inflation, the scalar field does not decay at the end of 
inflation because it needs to survive until today, to become quintessence.
This is why inflation is non-oscillatory (NO) and instead of the inflaton 
oscillating around its vacuum expectation value (VEV), it rolls down to the
quintessential plateau. Thus, the Universe must be reheated via a mechanism 
other than the decay of the inflaton field. 

A promising such reheating mechanism is instant preheating \cite{instant}, 
where, after inflation, the inflaton field crosses an enhanced symmetry point,
where it couples to some other field $\chi$. The non-adiabatic change of the 
$\chi$ effective mass results into copious production of $\chi$ particles, which
further decay into the thermal bath of the Hot Big Bang (HBB). Instant 
preheating can be very efficient, removing a large fraction of the inflaton's 
kinetic density. Another reheating mechanism for NO inflationary models is 
curvaton reheating \cite{curvreh}, where the Universe is reheated by the decay
of some spectator scalar field $\sigma$, that may or may not be responsible
for the curvature perturbation (if it is responsible it is called the curvaton).
The efficiency of curvaton reheating depends on the density budget of the 
curvaton at the time of its decay. The above mechanisms, however, introduce an 
additional field ($\chi$ or $\sigma$), which is to play a crucial role in the 
Universe history. As such, they are not aligned with the economy principle 
underlying quintessential inflation. 

Fortunately, there is another reheating mechanism, which does not rely on some
other scalar field playing a special role. This is the so-called gravitational 
reheating \cite{larry,ivonne}. Gravitational reheating is due to particle 
production during inflation of all light fields (i.e. with masses less than the 
Hubble scale), which are also non-conformally invariant. This is always present
in inflation, but the radiation density due to gravitational reheating is 
negligible in standard oscillatory inflation, so it is ignored. However, in NO 
inflation, this unavoidable radiation can be the only way to generate the 
thermal bath of the HBB.

As explained, the potential needs to have a huge drop in energy density 
between inflation and dark energy times. Soon after the end of inflation, as 
the potential energy undergoes this massive decrease, the scalar field becomes 
dominated by its kinetic density. If the latter also dominates over the 
background density, we enter a period of so-called kination \cite{kination}, 
until the Universe is reheated and the HBB begins. Soon after reheating, the 
field freezes and remains at a small constant potential density until much 
later, when it can play the role of quintessence.

Kination typically sends the field down to the quintessential plateau over 
a super-Planckian displacement in field space. This is a problem because 
radiative corrections threaten to lift the flatness of the potential.
Also, interactions with the standard model, albeit gravitationally suppressed,
may become important and challenge the equivalence principle. However, it is 
desirable that the field moves substantially down the potential, in order for
its potential density to massively decrease so that the gap between the energy 
scales of inflation and dark energy can be bridged. This is a \mbox{catch-22} 
problem of quintessential inflation; a super-Planckian displacement threatens 
the flatness of the quintessential plateau and may generate a fifth-force 
problem, but a sub-Planckian displacement makes it almost impossible to get 
from the inflationary to the quintessential plateau in such a way that the 
potential is not too curved during inflation, so that the generated curvature 
perturbation remains approximately scale-invariant (this is the $\eta$-problem 
of quintessential inflation \cite{eta}).

In this paper we attempt to address the above in the
context of $\alpha$-attractors in inflation model-building. The idea of 
$\alpha$-attractors is that the scalar field has a non-canonical kinetic term, 
which features poles. Such kinetic terms can be due to specific forms of the
K\"{a}hler potential in supergravity theories \cite{alpha}. The effect
of a pole in the kinetic term is that the field cannot travel through it in 
field space, so it imposes a bound on its value. Switching to a canonical field,
transposes the pole to infinity, while ``stretching'' the scalar potential of
the canonical field near the pole, generating thereby a plateau in the 
potential \cite{stretch}. Because of this, $\alpha$-attractors are rather 
popular for inflation model building \cite{restofalphas}, since the latest CMB 
data favour an inflationary plateau \cite{planck,BICEP2}. For quintessential 
inflation, we need two flat regions in the scalar potential and we show that 
this can be naturally generated within the standard $\alpha$-attractors 
framework.\footnote{See also Ref.~\cite{alphasami}, which appeared closely 
after our work. For pure quintessence with 
$\alpha$-attractors see Ref.~\cite{alphaQ}.} 
The ``stretching'' effect ensures that the 
plateaus in the potential are not in danger  from radiative corrections, even 
with a super-Planckian excursion of the canonical field because variation of 
the non-canonical field can be safely kept sub-Planckian by the bounds due to 
the poles in the kinetic term. As we explain, this also addresses the danger of 
the fifth force, so the above \mbox{catch-22} problem is overcome.\footnote{%
A super-Planckian excursion of the canonical field may result in the production
of sizeable gravitational waves, even though the variation of the non-canonical 
field is kept sub-Planckian. In this way, one can evade the Lyth bound and 
obtain a large value of the tensor-to-scalar ratio $r$ with a sub-Planckian 
(non-canonical) inflaton \cite{referee}. In our model, though, we only achieve 
a modest production of gravitational waves as we find  $r\lesssim 10^{-3}$.}

Our paper is structured as follows. In Sec.~2, we introduce our model. In 
Sec.~3, we discuss inflationary physics and obtain the inflationary observables,
such as the spectral index of the scalar curvature perturbations and the ratio 
between the spectra of tensor to scalar perturbations. We find that our model 
predictions fall very near the sweet spot of the latest CMB observations, as 
typical for a plateau inflation model. In Sec.~4, we study in detail the early 
history of the Universe after inflation. In particular we investigate, in a 
model independent way, kination and reheating, with emphasis on gravitational 
reheating. In Sec.~5, we discuss the physics of quintessence and constrain our 
model parameters such that the dark energy observations are satisfied. In 
Sec.~6, we discuss the problems of the fifth-force and overproduction of 
gravitinos and gravitational waves in our model. Finally, we conclude in Sec.~7.

We consider natural units, where $c=\hbar=1$ and Newton's gravitational 
constant is \mbox{$8\pi G=m_P^{-2}$}, with 
\mbox{$m_P\equiv M_P/\sqrt{8\pi}=2.43\times 10^{18}\,$GeV} 
being the reduced Planck mass.

\section{The Model}\leavevmode

We consider the following Lagrangian, 
\begin{equation}\label{eq:lagrangian1}
\mathscr{L} = \frac{\frac{1}{2}\partial_\mu\phi\,\partial^\mu\phi
}{\Big(1-\frac{\phi^2}{6\alpha}\Big)^2}\,m_P^2 - V_0e^{-\kappa\phi} + \Lambda	
\quad\,,
\end{equation}
where the dimensionless scalar field $\phi$ is measured in units of $m_P$,
$\alpha$ and $\kappa$ are dimensionless positive constants, $V_0$ is a constant
density scale and $\Lambda$ is the cosmological constant. In the above, the
non-canonical kinetic term of the field features poles at 
\mbox{$\phi=\pm\sqrt{6\alpha}$}, and has the standard form of $\alpha$-attractor
models motivated by supergravity \cite{alpha}, corresponding to a 
non-trivial K\"{a}hler manifold. In this context, the scalar potential can be 
due to non-perturbative effects, e.g. gaugino condensation \cite{gaugino}. The 
effect of the scalar potential is to drive $\phi$ to large values. However, the
existence of the poles in the kinetic term has the important consequence that 
the field cannot traverse through them in field space \cite{alpha}.

Quintessence was introduced to explain the dark energy observations 
\cite{review} without making use of the cosmological constant. 
The motivation is that the required value of the cosmological constant in
$\Lambda$CDM is incredibly fine-tuned because the vacuum density today is about 
$(10^{-3}\,{\rm eV})^4$. In our model we still feature $\Lambda$ but, as we show,
the required value is much more reasonable; at least as large as the 
electroweak scale. We introduce $\Lambda$ for the following reason.

As was standard 
practice until the observation of dark energy, we assume that, due to some 
unknown symmetry, the vacuum density in the Universe is zero. However, because 
of the positive pole present in the model, $\phi$ cannot go to infinity in the 
vacuum; it is capped at $\phi = \sqrt{6\alpha}$ because this is the value that 
corresponds to the smallest possible potential density. This means that zero 
vacuum energy density requires \mbox{$V_0e^{-\kappa\sqrt{6\alpha}}=\Lambda$}. 
Substituting this back into Eq.~\eqref{eq:lagrangian1}, the Lagrangian becomes

\begin{equation}\label{eq:lagrangian2}
\mathscr{L}=\frac{\frac{1}{2}(\partial\phi)^2m_P^2}%
{\Big(1-\frac{\phi^2}{6\alpha}\Big)^2} - 
V_0e^{-n}\left[e^{n\left(1-\frac{\phi}{\sqrt{6\alpha}}\right)}-1\right]	\quad\,,
\end{equation}
where $n\equiv\kappa\sqrt{6\alpha}$. It is now evident that as 
\mbox{$\phi\rightarrow\sqrt{6\alpha}$} the potential density disappears.
\footnote{One may contemplate adding an increment $\delta\Lambda$ to the value 
of the cosmological constant, which would appear in Eq.~\eqref{eq:lagrangian2}.
This would ensure eternal acceleration a la $\Lambda$CDM. In this case, 
considering the quintessential tail would only provide some dynamics to the 
effective barotropic parameter of dark energy, which might depart from $-1$. 
However, this increment $\delta\Lambda$ will suffer from the same problem as the
cosmological constant in $\Lambda$CDM, namely it would have to be incredibly 
fine-tuned so not to exceed the value of the dark energy density
at present $\sim(10^{-3}\,{\rm eV})^4$. In other words, the mechanism that is 
assumed to eliminate the vacuum density would have to deviate from exactly zero 
by this tiny amount. We feel that this would negate the need for quintessence 
and this is why we will not consider this possibility here.}

Now, the initial value of $\phi$ needs to be between the poles in the
potential. Were initially $\phi>\sqrt{6\alpha}$ then it would roll down to
infinity and the vacuum density would be zero without the introduction of 
$\Lambda$. Were initially $\phi<-\sqrt{6\alpha}$ then the field would roll 
down to $\phi=-\sqrt{6\alpha}$ only and the required $\Lambda$ for zero vacuum 
density would have been \mbox{$\Lambda=V_0e^{\kappa\sqrt{6\alpha}}$}. 
We do not consider either case. The reasons are practical. In the former case,
we have inflation near the pole but the exponential tail is not steep enough to
allow for successful quintessence. In the latter case, we have inflation with 
an exponential potential, which is power-law and contradicts observations (plus,
it never ends). Because of the no-hair theorem, the discussion over the intial 
conditions of the inflaton field is largely academic as it is not testable.
\footnote{Still a number of authors have considered the likelyhood of 
appropriate initial conditions for inflation (for a recent discussion see 
Ref.~\cite{bran}). One of the arguments is that the inflationary potential
must make contact with the Planck scale, otherwise the Universe cannot exit
from the spacetime foam and there is no initial boost for the expansion. The 
latest CMB observations favour plateau inflationary models, which cannot fulfill
this requirement since the potential density is capped at sub-Planckian values,
which may pose an initial condition problem for inflation \cite{stein}. One way 
to overcome this problem is by considering a period of power-law 
proto-inflation, which takes the system from the Planck scale and safely
places it at the inflationary plateau \cite{initial}.}

Because of the poles, the potential becomes
stretched at $\phi \rightarrow \pm \sqrt{6\alpha}$ \cite{stretch}. Hence, we 
find two plateaus in the model, with $V \rightarrow V_0e^{-n}(e^{2n}-1)$ or 
$V \rightarrow 0$. If the scalar field dominates the Universe, the two plateaus
can result in early and late periods of accelerated expansion. Thus, they can
play the role of the inflationary and quintessential plateau 
(also called `quintessential tail').

\begin{figure} [h]
\vspace{-8cm}

	\centering
	\includegraphics[width=\linewidth]{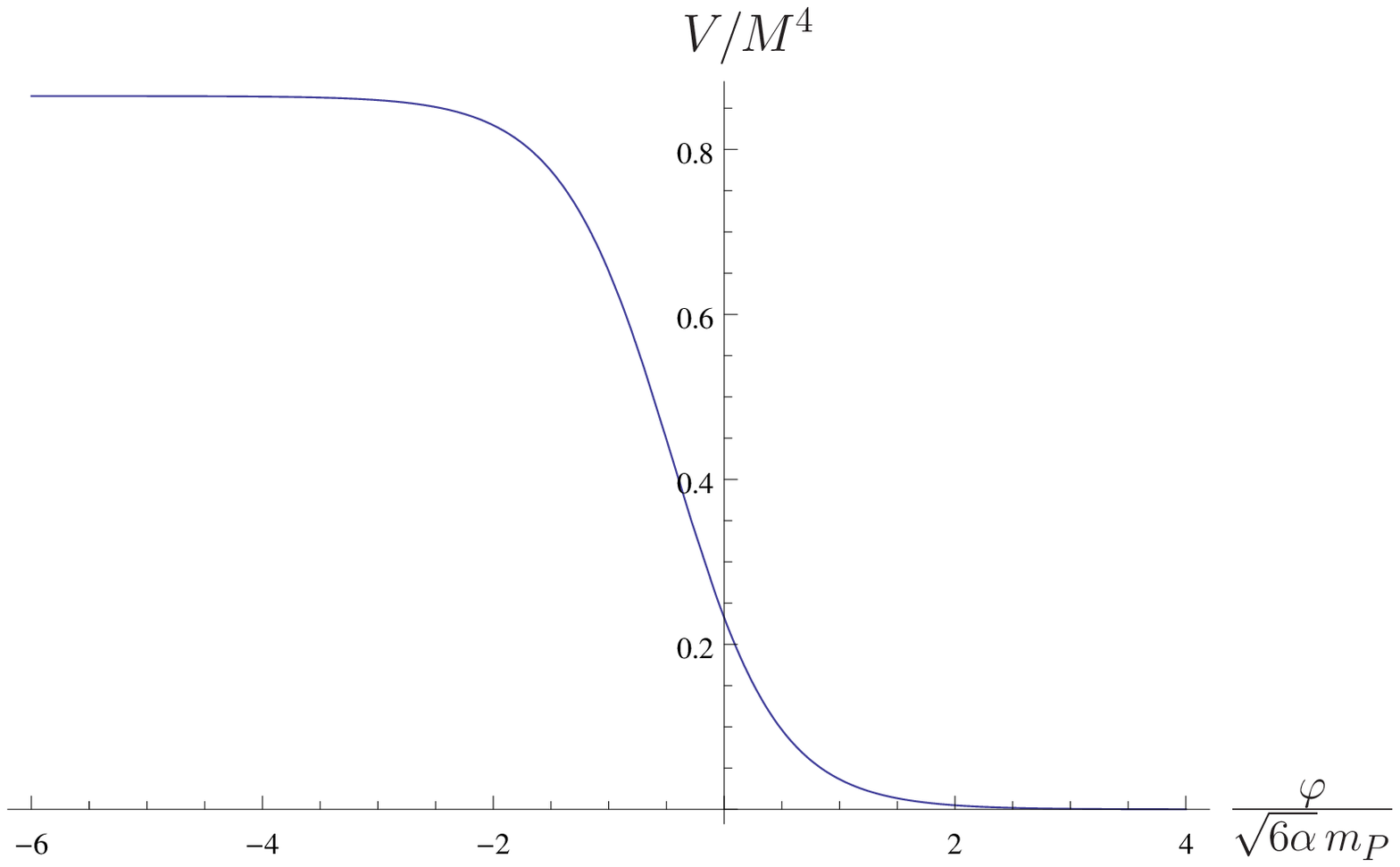} 
\vspace{-6cm}
	\caption{%
The potential in Eq.~\eqref{V2}. It features two flat regions for 
\mbox{$|\varphi|\gg\sqrt{6\alpha}\,m_P$}; the inflationary plateau and the 
quintessential tail, with a steep dip between them.}  
	\label{tanhpot} 
\end{figure}

To assist our intuition and help with studying the model,
we make the following field redefinition to obtain a canonical kinetic term
\cite{alpha}
\begin{equation}\label{eq:canonicalVariables}
\phi = \sqrt{6\alpha}\;\mathrm{tanh}\frac{\varphi}{\sqrt{6\alpha}m_P}	\quad\,,
\end{equation}
which allows our new canonical field, $\varphi$, to take any value whilst our 
non-canonical degree of freedom, $\phi$ remains sub-Planckian at all times, as 
long as $\alpha \lesssim \frac{1}{6}$. The potential becomes now:
%%%%
%%It is ok to use the non-canonical field for the sub-Planckian limit because this is the 'real' field, in the Lagrangian, the one that affects things, the one that has to be stable. However, all of the $n_s / r$ equations are derived presuming Einstein gravity and canonically normalised scalar fields - so by definition we must use the canonically normalised field for this - it is, after all, just a redefinition into the variables it wants.
%%%
\begin{equation}\label{V2}
V(\varphi) = e^{-2n}M^4 
\Big\{\exp\Big[n\Big(1-\tanh\frac{\varphi}{\sqrt{6\alpha}m_P}\Big)\Big]-1\Big\}	
\quad\,.
\end{equation}
where we have defined \mbox{$M^4 \equiv e^nV_0$}, which stands for the 
inflation energy scale. 
%because \mbox{$V\rightarrow (1-e^{-2n})M^4\simeq M^4$}
%when \mbox{$\varphi\rightarrow -\infty$} (we will find $n\gg 1$).
Note, also, that \mbox{$\Lambda=e^{-2n}M$}.
The potential is shown in Fig.~\ref{tanhpot}.

As $\mathrm{tanh}(\varphi/\sqrt{6\alpha}\,m_P)$ approaches a constant value 
when $|\varphi|$ is very large, the potential becomes asymptotically constant,
featuring plateaus. At the locations of these plateaus the field slow rolls and
accelerated expansion occurs. In the following sections, 
we examine these two periods; that of inflation when
$\varphi\rightarrow-\infty$ ($\phi \rightarrow -\sqrt{6\alpha}$) and 
that of  quintessence when
$\varphi\rightarrow +\infty$ ($\phi \rightarrow \sqrt{6\alpha}$), 
as well as the evolution between them.

\section{Inflation}\leavevmode

In the limit $\varphi\rightarrow -\infty$ ($\phi \rightarrow -\sqrt{6\alpha}$), 
the potential in Eq.~\eqref{V2} becomes:

\begin{equation}\label{eq:Infl_pot}
V(\varphi)\simeq M^4\mathrm{exp}\Big(-2ne^{\frac{2\varphi}{\sqrt{6\alpha}m_P}}\Big)	
\quad\,.
\end{equation}

\subsection%{\boldmath $n_s$ and $r$}
{The Scalar Spectral Index and Tensor Ratio}
\label{nsandr}

In view of the above, the slow roll parameters are

\begin{equation}\label{eq:epsilon_phi}
\epsilon = \frac{m_P^2}{2}\Big(\frac{V'}{V}\Big)^2 = 
\frac{4n^2}{3\alpha}\,e^{\frac{4\varphi_*}{\sqrt{6\alpha}m_P}}	\quad\,,
\end{equation}

\begin{equation}\label{eq:eta_phi}
\eta = m_P^2 \frac{V''}{V} = 
-\frac{4n}{3\alpha}\,e^{\frac{2\varphi_*}{\sqrt{6\alpha}m_P}} 
\left(1 - 2n e^{\frac{2\varphi_*}{\sqrt{6\alpha}m_P}}\right)	\quad\,,
\end{equation}
where `$*$' denotes the value at horizon crossing, when cosmological scales 
exit the horizon and the prime denotes derivative with respect to $\varphi$. 
From the usual condition of the end of inflation, 
$\epsilon = 1$, we find:
	
\begin{equation}\label{eq:phi_e}
\varphi_{\mathrm{end}} = 
\frac{\sqrt{6\alpha}}{2}\,m_P\,
\mathrm{ln}\Big(\frac{\sqrt{3\alpha}}{2n}\Big)	\quad\,.
\end{equation}
The slow roll parameters are better expressed as functions of the number of 
remaining e-folds of inflation at horizon crossing of cosmological scales, 
defined as:

\begin{equation}\label{e-folds_main}
N_*=\frac{1}{m_P^2}\int_{\varphi_\mathrm{end}}^{\varphi_*}\frac{V}{V'}
\,\mathrm{d}\varphi
	\quad\,,
\end{equation}
through which we obtain:

\begin{equation}\label{eq:phi_N}
\varphi_* = \frac{\sqrt{6\alpha}}{2}\,m_P\, 
\mathrm{ln}\left[\frac{3\alpha}{4n}\Big(N_*+\frac{\sqrt{3\alpha}}{2}\Big)^{-1}
\right]	\quad\,,
\end{equation}
which can be negative if $\alpha$ is small. Using the above,
%This allows us to eliminate $\varphi_*$ in 
the slow-roll parameters
%and instead replace it with $N_*$ to find:
become
\begin{equation}\label{eq:epsilon_N}
\epsilon = \frac{3\alpha}{4}\Big(N_* + \frac{\sqrt{3\alpha}}{2}\Big)^{-2}	
\quad\,,
\end{equation}

\begin{equation}\label{eq:eta_N}
\eta = -\Big(N_* + \frac{\sqrt{3\alpha}}{2}\Big)^{-1}
\left[1-\frac{3\alpha}{2}\Big(N_*+\frac{\sqrt{3\alpha}}{2}\Big)^{-1}\right]	
\quad\,.
\end{equation}
Thus, we obtain the tensor-to-scalar ratio and the spectral index of the scalar 
curvature perturbation as %, in terms of e-folding number:

\begin{equation}\label{eq:r}
r = 16\epsilon = 12\alpha\Big(N_* + \frac{\sqrt{3\alpha}}{2}\Big)^{-2}	\quad\,,
\end{equation}

\begin{equation}\label{eq:n_s_final}
n_s=1+2\eta-6\epsilon=1-\frac{2}{\Big(N_*+\frac{\sqrt{3\alpha}}{2}\Big)} - 
\frac{3\alpha}{2\Big(N_* + \frac{\sqrt{3\alpha}}{2}\Big)^2} 
%\qquad 
\;\simeq\; 
%\qquad 
1 - \frac{2}{N_*}	%\quad	\mathrm{for \; small } \; \alpha\quad\,.
\end{equation}
where the last equation in Eq.~\eqref{eq:n_s_final} corresponds to small 
$\alpha$. We see that $n_s$ follows the pattern of the $\alpha$-attractors 
inflationary models \cite{alpha,restofalphas}. In fact, this is suggested by 
most plateau inflationary models (e.g. see Ref.~\cite{ours}), like Starobinsky
\cite{R2} and Higgs \cite{Higgs} inflation, which are favoured by the latest 
CMB observations \cite{planck,BICEP2}.
%The T-model $\alpha$-attractors models have potentials of the form $V \simeq V_0\mathrm{tanh}\frac{\varphi}{\sqrt{6\alpha}}$ and different values of alpha correspond to sometimes recognisable models of inflation. For example are both members of the $\alpha-$attractors family. It will be interesting to see if our final constrained values of $\alpha$ align with anything recognisable in quintessential inflation.

From the above, the running of the spectral index is easy to calculate as
\begin{equation}
n_s'\equiv\frac{{\rm d}\ln n_s}{{\rm d}\ln k}=
-\frac{1}{\Big(N_*+\frac{\sqrt{3\alpha}}{2}\Big)}
\frac{2\Big(N_*+\frac{\sqrt{3\alpha}}{2}\Big)+3\alpha}
{\Big(N_*+\frac{\sqrt{3\alpha}}{2}\Big)^2-
2\Big(N_*+\frac{\sqrt{3\alpha}}{2}\Big)-\frac32\alpha}\simeq
-\frac{2}{N_*^2-2N_*}
\label{nsrunning}
\end{equation}
where again the last equation in the above corresponds to small $\alpha$.

Here we should briefly consider what it really means when we apply the limits
\mbox{$\alpha\rightarrow 0$} and \mbox{$\alpha\rightarrow\infty$}. If 
%\mbox{$\alpha\ll 1$}
\mbox{$\alpha\rightarrow 0$} 
then the region between the poles is shrinking, 
so it becomes increasingly unlikely that $\phi$ initially finds itself there.
Still, as we show below, when \mbox{$\alpha\lesssim 0.1$} or so the value of
the spectral index gradually becomes insensitive to $\alpha$ (see 
Fig.~\ref{fig:n_s_and_r}), which means that
there is no point considering $\alpha$ incredibly small (which would amount 
to fine-tuning anyway). In the opposite limit, \mbox{$\alpha\rightarrow\infty$},
the poles are transposed to infinity and $\phi$ becomes canonically normalised.
In this case, there are no plateaus to consider and we end up with either 
power-law  inflation that never ends, or with no inflation at all (depending on
how big $\kappa$ is in Eq.~\eqref{eq:lagrangian1}). Barring the extremes 
\mbox{$\alpha=0,\infty$}, the natural value of $\alpha$ is close to unity.

\subsection{\boldmath Constraining $N_*$ and $M$}
The number of remaining e-folds of inflation when the cosmological scales exit 
the horizon, $N_*$ depends on the expansion history of the Universe. In this 
model we have a period of kination, where the kinetic energy density of the 
inflaton is, for a time, the dominant energy density in the Universe and 
controls its evolution. During this regime $a \propto \rho^{-1/6}$. 

We start from the recognisable equation

\begin{equation}%\label{key}
e^{N_*} = 2\frac{H_*}{H_k}\Big(\frac{a_{\mathrm{end}}}{a_{\mathrm{reh}}}\Big)
\Big(\frac{a_{\mathrm{reh}}}{a_{\mathrm{eq}}}\Big)
\Big(\frac{a_{\mathrm{eq}}}{a_k}\Big)	\qquad \,,
\end{equation}
where subscripts `end', `reh', and `eq' refer to the end of inflation, the 
onset of radiation domination and the beginning of matter domination 
respectively, while subscript `$k$' corresponds to horizon reentry of the 
pivot scale $k = 0.05\,\mathrm{Mpc^{-1}}$. From the above, we obtain

\begin{equation}\label{eq:Nstar}
N_*\simeq 61.93 + \mathrm{ln}\Big(\frac{V_{\mathrm{end}}^{1/4}}{m_P}\Big) + 
\frac{1}{3}\mathrm{ln}\Big(\frac{V_{\mathrm{end}}^{1/4}}{T_{\mathrm{reh}}}\Big) 	
\qquad \,,
\end{equation}
where $T_{\rm reh}$ is the reheating temperature when 
the Hot Big Bang (HBB) begins and %\\ 
\mbox{$V_{\rm end}\equiv V(\varphi_{\mathrm{end}})$}
$=M^4e^{-\sqrt{3\alpha}}$ 
(cf. Eq.~\eqref{eq:phi_e}). This differs slightly from that of a model which 
contains no kination \cite{Mukhanov:2005sc}. 
%Derived fully in Section \ref{reh}, the reheating temperature is:
As shown in Sec.~\ref{reh} (cf. Eq.~\eqref{eq:T_reh}), for gravitational 
reheating \mbox{$T_{\rm reh}\propto V_{\rm end}/m_P^3$}.
%\begin{equation}
%T_{\rm reh}=\frac{q^{3/4}}{36\pi^2}
%\left(\frac{g_*^{\rm end}}{g_*^{\rm reh}}\right)^{1/4}
%\sqrt{\frac{g_*^{\rm end}}{10}}\,\frac{V_\mathrm{end}}{m_P^3}\quad.
%\end{equation}
%where $q$ is a reheating efficiency factor and $g_{*}^{\mathrm{end}}(g_{*}^{\mathrm{reh}})$ are the number of effective relativistic degrees of freedom at the end of inflation and kination, respectively. 
%After some brief algebra we find 
Using this, it is easy to show that
\mbox{$\big(\frac{V_{\mathrm{end}}^{1/4}}{T_{\mathrm{reh}}}\big)^{^{1/3}}\!\!\propto 
\big(\frac{m_P}{V_{\mathrm{end}}^{1/4}}\big)$}. 
%, where $k$ is a constant, and so 
Thus, the dependence on $V_{\mathrm{end}}$ in the second and third terms of 
Eq.~\eqref{eq:Nstar} cancels out and we are left with a constant value for 
$N_*$, independent of both $\alpha$ and $n$:

\begin{equation}\label{eq:final_N_star}
N_{*} = 63.49%3	
\qquad \,.
\end{equation}
Using the above, Eqs.~\eqref{eq:n_s_final} and \eqref{nsrunning} give
\mbox{$n_s\simeq 0.9685$} and \mbox{$n_s'\simeq-5.11\times 10^{-4}$} for 
negligible $\alpha$, which is in excellent agreement with observations
(\mbox{$n_s=0.968\pm0.006$} and \mbox{$n_s'=-0.003\pm0.007$}~\cite{planck}).

We can determine the inflationary scale by the so-called COBE 
constraint~\cite{book}

\begin{equation}\label{eq:GenericInflationaryScale}
\sqrt{\mathscr{P}_{\zeta}} = \frac{1}{2\sqrt{3}\pi}\frac{V^{3/2}}{m_P^3 |V'|}	
\qquad \,,
\end{equation}
where $\mathscr{P}_{\zeta} = (2.208 \pm 0.075) \times 10^{-9}$, is the spectrum 
of the scalar curvature perturbation \cite{planck}. Using 
Eqs.~\eqref{eq:Infl_pot} and \eqref{eq:phi_N} we find:

\begin{equation}\label{eq:InflationaryScale}
\left(\frac{M}{m_P}\right)^2= 3\pi\sqrt{2\alpha\mathscr{P}_{\zeta}}
\Big(N_*+\frac{\sqrt{3\alpha}}{2}\Big)^{-1}
\mathrm{exp}
\left[\frac{3\alpha}{4}\Big(N_*+\frac{\sqrt{3\alpha}}{2}\Big)^{-1}\right]
\qquad	\,.
\end{equation}
We determine the particular values of $M$ for various $\alpha$ values, shown in 
Table~\ref{table:Nforalpha}. Eq.~\eqref{eq:InflationaryScale} suggests that the 
inflation energy scale $M$ is also independent of $n$. As expected, $M$ is near 
the scale of grand unification.

\begin{table}[h]
	\begin{center}
		\begin{tabular}{|c|c|}
\hline $\alpha$    & $M$(GeV) \\ 
\hline 0.01        & $2.42\times 10^{15}$ \\ 
%\hline 0.05        & $3.61\times 10^{15}$ \\ 
\hline 0.10        & $4.29\times 10^{15}$ \\ 
%\hline 0.50        & $6.42\times 10^{15}$ \\ 
\hline 1           & $7.64\times 10^{15}$ \\ 
%\hline 5           & $1.16\times 10^{16}$ \\ 
\hline 10          & $1.41\times 10^{16}$ \\ 
%\hline 50          & $2.54\times 10^{16}$ \\ 
\hline 100         & $3.81\times 10^{16}$ \\ 
\hline 
		\end{tabular}  
	\end{center}
\caption{Values of $M$ calculated from Eq~\eqref{eq:InflationaryScale} 
for various $\alpha$ values.}
	\label{table:Nforalpha}
\end{table}

\subsection{Parameter Space from Observational Constraints}
\leavevmode
We are able to test the constraints of the model immediately via comparison of 
the model's prediction for the tensor-to-scalar ratio to observation. The 
constraint on the tensor-to-scalar ratio, $r < 0.1$~\cite{Ade:2015lrj}, allows us to constrain the allowed values of $\alpha$:

\begin{equation}%\label{key}
120\,\alpha < \Big(N_* + \frac{\sqrt{3\alpha}}{2}\Big)^{2}	\qquad\,.
\end{equation}
Using the value of $N_*$ derived in the previous section, $N_{*} = 63.49$, 
results in a bound of $\alpha \leq 39.6$. Currently 
we have no lower bound on $r$ and hence no lower bound on $\alpha$, but (as 
shown later) it should not get too small. Requiring the mass scale which 
suppresses the non-canonical $\phi$ in the kinetic term not to be too small 
compared to $M$, we have \mbox{$\sqrt{6\alpha}\,m_P \gtrsim M$}, which results in \mbox{$\alpha\gtrsim 10 ^{-7}$}. 

\begin{figure} [h]
	\centering
	\includegraphics[width=.6\linewidth]{%Graph_1.png
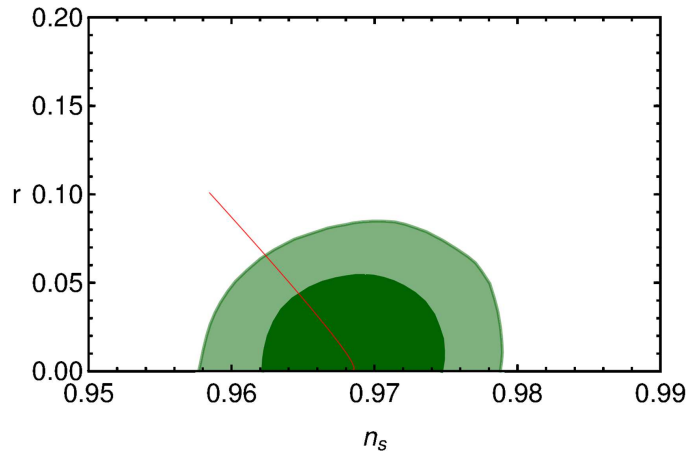} 
	\caption{The tensor-to-scalar ratio, $r$, versus the spectral index, 
$n_s$ for our model is displayed in red, overlaid on the Planck 2015 results. 
$\alpha$ varies from 0 to 39.6, bottom to top. The slope of the line for large 
values of $\alpha$ is understood as \mbox{$n_s\rightarrow 0$} when
\mbox{$\alpha\gg 1$} (cf. Eq.~\eqref{eq:n_s_final}).
Note that the line corresponding to the values of $n_s$ and $r$ is crooked for 
small $\alpha$ (values of \mbox{$\alpha\lesssim 0.1$} or so) so that the 
spectral index becomes insensitive to $\alpha$ when the latter is small. For a 
closer look at this region see Fig.~\ref{fig:n_s_and_r2}.}
	\label{fig:n_s_and_r} 
\end{figure}

The corresponding bounds on $n_s$ are

\begin{equation}%\label{key}
0.9585 \lesssim n_s \leq 0.9686		\qquad \,,
\end{equation}
and fall almost entirely within the recent BICEP2 2-$\sigma$ bounds 
\cite{BICEP2}. Fig.~\ref{fig:n_s_and_r} shows the parameter space for $n_s$ and 
$r$ in our model, using these values. However, because of the field redefinition
in Eq.~\eqref{eq:canonicalVariables}, to avoid super-Planckian values of $\phi$,
we need $\alpha\lesssim\frac16$, which places us firmly within the constraints 
of the tensor-to-scalar ratio at $r \leq 0.00049$. The bounds on $n_s$ with this
$\alpha$ constraint are:

\begin{equation}%\label{key}
0.9685 \lesssim n_s \leq  0.9686	\qquad \,,
\end{equation}
which corresponds to the very lowest part of the line in Fig.~\ref{fig:n_s_and_r}, well inside the Planck 1-$\sigma$ contour.

%%%%%%%%%%%%%%%%%%
% DESCRIPTION OF MODEL IN BETWEEN THE TWO REGIMES
%%%%%%%%%%%%%%%%%%

%\clearpage

\section{After Inflation}
%Between Inflation and Late-time Acceleration}

During inflation the inflaton is slow rolling along the inflationary plateau
but, after the end of inflation, the inflaton falls down the steep slope of the 
potential. For a brief period of time, the kinetic energy of the inflaton is the
dominant energy density in the Universe, the field is oblivious to the potential
and we have a period of so-called kination \cite{kination}. Kination, however, 
must end well before Big Bang Nucleosynthesis (BBN). Hence, the Universe should
be reheated so that the HBB can begin. Because inflation is non-oscillatory, 
reheating must occur without the decay of the inflaton field, which is to
survive until the present and become quintessence. In the following, we will 
briefly study kination and reheating.

\subsection{Kination}

Soon after inflation ends, the inflaton energy density is completely dominated 
by the kinetic part as the potential density becomes negligible. Being 
oblivious of the potential, the equation of motion becomes
\mbox{$\ddot\varphi+3H\dot\varphi\simeq 0$}, which gives
\begin{equation}\label{eq:phikin}
\dot{\varphi} = \sqrt{\frac{2}{3}}\frac{m_P}{t}\quad.
\end{equation}
Thus, in kination we have
\mbox{$\rho=\rho_{\rm kin}\equiv\frac12\dot\varphi^2\propto a^{-6}$} with 
$a\propto t^{1/3}$ and barotropic parameter $w=1$ (stiff fluid). 
Integrating Eq.~\eqref{eq:phikin} we obtain

\begin{equation}\label{eq:phi_*_2}
\varphi=\varphi_{\rm end}+\sqrt{\frac23}\,m_P\ln\left(\frac{t}{t_{\rm end}}\right)
\qquad \,.
\end{equation}
Kination has to end and the HBB to begin before BBN 
takes place. Therefore, the radiation bath of the HBB must be created after 
inflation. Because radiation density scales as \mbox{$\rho_\gamma\propto a^{-4}$},
once created, radiation eventually takes over, since 
\mbox{$\rho_{\rm kin}\propto a^{-6}$}, but 
this has to happen before BBN. Also, since the inflaton is not oscillating after
inflation, reheating of the Universe cannot occur through the %perturbative 
decay of the inflaton field.

\subsection{Reheating}
\label{reh}

There are various possibilities for reheating the Universe in quintessential
inflation scenarios, most promising of which are
instant preheating \cite{instant} and curvaton reheating \cite{curvreh}. If 
all else fails, the Universe is reheated by so-called gravitational reheating
\cite{larry,ivonne}. As mentioned, 
the latter is due to particle production during inflation of all light fields
(i.e. with masses less than the Hubble scale), which are also non-conformally
invariant. This is Hawking 
radiation in de Sitter space, which generates a radiation bath of temperature 
\mbox{$T_H=H/2\pi$}. Such radiation is always produced at the end of inflation, 
but its density $\sim T_H^4$ is negligible in standard oscillatory inflation, 
so it is ignored. 

The radiation density produced through gravitational reheating is
\begin{equation}
(\rho_\gamma^{\rm gr})_{\rm end}=q\,
\frac{\pi^2}{30}g_*^{\rm end}\left(\frac{H_{\rm end}}{2\pi}\right)^4=
\frac{q\,g_*^{\rm end}}{480\pi^2}\,H_{\rm end}^4\quad,
\label{eq: rhograv}
\end{equation}
where $g_*^{\rm end}={\cal O}(100)$ is the number of effective relativistic 
degrees of freedom at the energy scale of inflation and \mbox{$q\sim 1$} is 
some efficiency factor. The above does not imply that radiation produced through
gravitational reheating is thermal. Indeed, even though it has been found 
that \mbox{$(\rho_\gamma^{\rm gr})_{\rm end}\sim 10^{-2}H_{\rm end}^4$} 
\cite{larry,ivonne}, as suggested above, thermalisation of the produced 
radiation may occur much later \cite{ivonne}. This, however, does not make any 
difference in our considerations, because radiation always scales as 
$\rho_\gamma\propto a^{-4}$ regardless of whether it is thermalised or not. 

In view of Eq.~\eqref{eq: rhograv},
the density parameter of radiation for gravitational 
reheating at the end of inflation is 
\begin{equation}
(\Omega_\gamma^{\rm end})_{\rm gr}\equiv
\left.\frac{\rho_\gamma^{\rm gr}}{\rho}\right|_{\rm end}=
\frac{q\,g_*^{\rm end}}{1440\pi^2}\left(\frac{H_{\rm end}}{m_P}\right)^2
\quad.
\label{greff}
\end{equation}
The above is the lowest possible value of the radiation density parameter
at the end of inflation 
\mbox{$\Omega_\gamma^{\rm end}\equiv(\rho_\gamma/\rho)_{\rm end}$}, which can, 
in principle 
approach unity, in the case of instant preheating. Thus, in general we have
\begin{equation}
(\Omega_\gamma^{\rm end})_{\rm gr}\lesssim\Omega_\gamma^{\rm end}\lesssim 1\quad.
\label{Orange}
\end{equation}

The Universe is reheated when the radiation takes over and dominates the 
kinetic density of the scalar field. This is bound to happen, regardless how 
small $\Omega_\gamma^{\rm end}$ is, because 
\mbox{$\rho_{\rm kin}\propto a^{-6}$}, while for radiation we have 
\mbox{$\rho_\gamma\propto a^{-4}$}.
Using that $a\propto t^{1/3}$ during kination,
it is straightforward to show that the time when the HBB begins 
(i.e. radiation takes over) is
\begin{equation}
t_{\rm reh}=(\Omega_\gamma^{\rm end})^{-3/2}t_{\rm end}
\end{equation}
Then Eq.~\eqref{eq:phi_*_2} gives
\begin{equation}
\varphi_{\rm reh}=
\varphi_{\rm end}-\mbox{$\sqrt{\frac32}$}\,m_P\ln\Omega_\gamma^{\rm end}\quad.
\quad.
\label{phi_kin}
\end{equation}

Now, considering that radiation is thermalised by the time it comes to
dominate the Universe (this is certainly true for gravitational reheating,
where \ $\Omega_\gamma$ is minimal), the reheating temperature is obtained as 
follows. Since \mbox{$\Omega_\gamma=\rho_\gamma/\rho_{\rm kin}\propto a^2$} during
kination, it is easy to find that 
\mbox{$\rho_{\rm kin}^{\rm reh}=(\Omega_\gamma^{\rm end})^3\rho_\phi^{\rm end}$},
where $\rho_{\rm kin}^{\rm reh}\equiv\rho_{\rm kin}(t_{\rm reh})$. 
Using that \mbox{$\rho_\gamma^{\rm reh}\equiv\rho_{\rm kin}^{\rm reh}$},
the reheating temperature is
\begin{equation}
T_{\rm reh}=\left[
\frac{30}{\pi^2 g_*^{\rm reh}}(\Omega_\gamma^{\rm end})^3\rho_\phi^{\rm end}
\right]^{1/4}\quad.
%\propto (\Omega_\gamma^{\rm end})^{3/4}
\end{equation}
Combining the above with Eqs.~(\ref{greff}) and (\ref{Orange}), we find
\begin{equation}\label{Treh}
T_{\rm reh}\geq\frac{q^{3/4}}{24\pi^2}
\left(\frac{g_*^{\rm end}}{g_*^{\rm reh}}\right)^{1/4}
\sqrt{\frac{g_*^{\rm end}}{10}}\,\frac{H_{\rm end}^2}{m_P}\quad,
\end{equation}
where the equality corresponds to gravitational reheating. For inflation near 
the grand unified energy scale (cf. Table~\ref{table:Nforalpha}) we have
\mbox{$H_{\rm end}\sim 10^{12}\,$GeV}.
Taking %\mbox{$g_*^{\rm end}=106.75$} 
\mbox{$g_*^{\rm end}={\cal O}(100)$} 
and \mbox{$g_*^{\rm reh}=10.75$} (assuming 
late reheating), we find \mbox{$T_{\rm reh}\gtrsim 10^4\,$GeV}, which
is safely much higher than the temperature at BBN.\footnote{Quintessential 
inflation typically features low values of $T_{\rm reh}$. For example, 
\mbox{$T_{\rm reh}\sim 10^4\,$GeV} is an {\em upper} bound in the particular 
quintessential inflation model in Ref.~\cite{QIreh}.}

\subsection{Freezing of the Scalar Field}

\begin{figure} [h]
%\vspace{-8cm}

	\centering
	\includegraphics[width=.8\linewidth]{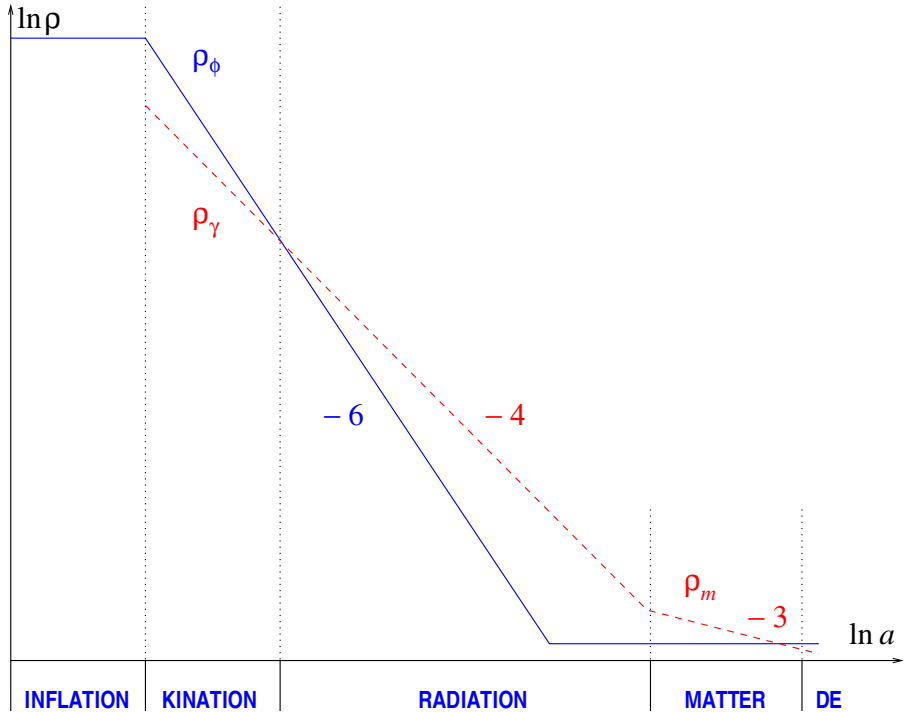} 
%\vspace{-6cm}
	\caption{%
Log-log plot depicting the evolution of the density of the scalar field
$\rho_\varphi$ (solid line) and the radiation density of the HBB $\rho_\gamma$, 
which eventually gives away to matter $\rho_m$ (both depicted by the dashed 
line). In late times the Universe is dominated by dark energy (DE in the graph).
}  
	\label{kination} 
\end{figure}

After kination ends and radiation domination takes over, the field continues to
roll for a time until it runs out of kinetic energy and freezes. 
Indeed, after the onset of the HBB, the field is still kinetically dominated
so that the equation of motion is still 
\mbox{$\ddot\varphi+3H\dot\varphi\simeq 0$}.
However, in radiation domination, this results in 
\begin{equation}\label{eq:dot_phi_RD}
\dot{\varphi} = \sqrt{\frac23}\frac{m_P\sqrt{t_{\rm reh}}}{t^{3/2}}	
\qquad \,.
\end{equation}
Integrating the above we find
\begin{equation}
\varphi=\varphi_{\rm reh}+
\sqrt{\frac23}\,m_P\left(1-\sqrt{\frac{t_{\rm reh}}{t}}\right)\quad.
\label{phiRD}
\end{equation}
Thus, the field freezes for \mbox{$t\gg t_{\rm reh}$} at the value
\begin{equation}
\varphi_F=\varphi_{\rm end}+\mbox{$\sqrt{\frac23}$}
\left(1-\mbox{$\frac32$}\ln\Omega_\gamma^{\rm end}\right)m_P\quad,
\label{eq:phi_f_3}
\end{equation}
where we considered also Eq.~\eqref{phi_kin}. It should be stressed here that
this result is model independent because, while $\varphi$ is kinetically
dominated, it is oblivious to the potential.\footnote{Of course, just before
freezing we have \mbox{$\rho_{\rm kin}\lesssim V$}. But the subsequent variation
of $\varphi$ is exponentially suppressed, so \mbox{$\varphi\simeq\varphi_F=\,$
constant}.} The evolution of $\rho_\varphi$ is shown in Fig.~\ref{kination}.

From Eq.~\eqref{eq:phi_f_3}, to maximise the value of 
$\varphi_F$, in order to achieve a low residual potential density, we see that 
we have to consider the minimum possible value of $\Omega_\gamma$. As suggested
by Eq.~\eqref{Orange}, this corresponds to gravitational reheating.

Therefore, in this paper we consider gravitational reheating. As explained, 
gravitational reheating is selected to ensure the field retains enough kinetic 
density to roll to a low potential density, to align with observations of dark 
energy today. Additionally, this negates the need to introduce additional scalar
fields that play a crucial role in reheating, as in the cases of instant 
preheating and curvaton reheating. This promotes economy in the
model because it ensures our model stays as minimal as possible. 

%%%%%%%%%%%%%%%%%%%%%%%%%%%%%%%%%%%%%%%%%%%%
%QUINTESSENCE REGIME
%%%%%%%%%%%%%%%%%%%%%%%%%%%%%%%%%%%%%%%%%%%%

\section{Quintessence}\leavevmode

%Investigating the regime of quintessence, we are i
In the limit $\varphi\rightarrow+\infty$ ($\phi \rightarrow \sqrt{6\alpha}$),
%. Starting again from 
the potential in Eq. \eqref{V2} becomes
%, with $\mathrm{tanh}$ in its exponential form and simplifying for the large field regime we find:

\begin{equation}\label{eq:Quint_V}
V \simeq 2ne^{-2n}M^4 e^{-\frac{2\varphi}{\sqrt{6\alpha}m_P}} \quad\,.
\end{equation}
Thus, the potential features a classic exponential quintessential tail, of the 
form 

\begin{equation}\label{eq:V_atr_1}
V=V_{\mathrm{Q}}\exp\left(-\lambda\varphi/m_P\right)\quad,
\end{equation}
where $V_Q=2ne^{-2n}M^4$ and $\lambda=2/\sqrt{6\alpha}=(2/n)\kappa$. The 
exponential quintessential tail features two attractor solutions, depending
on whether the scalar field is dominant or not to the background density.
Originally, the field is frozen at $\varphi_F$, in Eq.~\eqref{eq:phi_f_3},
with potential density \mbox{$V_F\equiv V(\varphi_F)$}. However, when $V_F$
approaches the potential density
of the attractor, the field unfreezes and eventually follows the attractor 
solution. As shown in numerical simulations \cite{Original_Trackers_Quint},
the system briefly oscillates around the attractor before following it.
Below we consider both attractors and the corresponding parameter space.
It is easy to check that the attractor solutions in Eqs.~\eqref{att1} and
\eqref{att2} are solutions to the Klein-Gordon equation for the canonical
field
\begin{equation}
\ddot\varphi+3H\dot\varphi+V'=0\quad.
\end{equation}
\subsection{Dominant Quintessence}

If the scalar field is dominant then the attractor is
\begin{equation}
V=%p(3p-1)
\frac{2(6-\lambda^2)}{\lambda^4}\left(\frac{m_P}{t}\right)^2\quad\&\quad
\rho_{\rm kin}\equiv
\frac12\dot\varphi^2=\frac{2}{\lambda^2}\left(\frac{m_P}{t}\right)^2
\quad\Rightarrow\quad
\rho_\varphi=\frac{12}{\lambda^4}\left(\frac{m_P}{t}\right)^2\quad,
\label{att1}
\end{equation}
where \mbox{$\rho_\varphi=\frac12\dot\varphi^2+V$}. Then, 
%it is easy to show that
\mbox{$a\propto t^{2/\lambda^2}=t^{1/\epsilon}\Rightarrow H\propto a^{-\epsilon}$},
where \mbox{$\epsilon\equiv-\dot H/H^2=\lambda^2/2$}. Since,
\mbox{$a\propto t^{2/3(1+w)}$} we find that the barotropic parameter of the 
Universe is \mbox{$w=-1+\lambda^2/3$}. Thus, in order to have accelerated 
expansion, we require \mbox{$w<-\frac13$}, i.e. \mbox{$\lambda<\sqrt 2$}.
This accelerated expansion is eternal. The Planck constraint on the barotropic
parameter of dark energy is \mbox{$w=-1.006\pm0.045$} \cite{planck}.
This means that \mbox{$\lambda\leq 0.342$}, which results in the bound
\mbox{$\sqrt{6\alpha}=2/\lambda\geq 5.847$}. Such a large $\alpha$ would render 
$\phi$ super-Planckian, which should be avoided.

Alternatively, for \mbox{$\sqrt 2\leq\lambda<\sqrt 3$} we still have a
negative pressure but it is not negative enough to lead to eternal accelerated 
expansion. It can lead, however, to transient accelerated expansion 
\cite{trans1}. Indeed, as mentioned above, after the field unfreezes, the system
briefly oscillates around the attractor. As such, the effective barotropic 
parameter can temporarily decrease below $-\frac13$ so expansion becomes 
accelerated.

\subsection{Subdominant Quintessence}

If the scalar field is subdominant then the attractor is
\begin{equation}
V=\frac{2}{\lambda^2}\left(\frac{1-w}{1+w}\right)
\left(\frac{m_P}{t}\right)^2\quad\&\quad
\rho_{\rm kin}\equiv
\frac12\dot\varphi^2=\frac{2}{\lambda^2}\left(\frac{m_P}{t}\right)^2
\quad\Rightarrow\quad
\rho_\varphi=\frac{4}{\lambda^2(1+w)}\left(\frac{m_P}{t}\right)^2\quad,
\label{att2}
\end{equation}
where $w$ now is the barotropic parameter of the background matter 
($w=0$ in the case of matter). We know 
\begin{equation}%\label{key}
\rho = \frac{4}{3(1+w)^2}\left(\frac{m_P}{t}\right)^2	\quad.
\end{equation}
So we find

\begin{equation}%\label{key}
\Omega_{\varphi}\equiv\frac{\rho_\varphi}{\rho}=\frac{3(1+w)}{\lambda^2}\quad.
\end{equation}
For subdominant quintessence, $\Omega_{\varphi}<1$, which implies
\mbox{$\lambda>\sqrt{3(1+w)}$}.

The attractor in the subdominant quintessence case does not lead to 
accelerated expansion, because the evolution of the density of the scalar field
mimics the background density. However, similarly to the previous case,
because, after unfreezing, the system briefly oscillates around the attractor, 
this may result into a bout of transient accelerated expansion, as the scalar 
field density briefly dominates before settling to its path, just below the 
background density. For this, the quintessence density should not be much 
smaller than the background density. Numerical studies have shown that this
occurs for $\lambda^2<24$ or so \cite{eta,trans2}.

Thus, for transient accelerated expansion we need
\begin{equation}
\sqrt 2\leq\lambda\lesssim 2\sqrt 6\quad,
\label{transient}
\end{equation}
while $\lambda<\sqrt 2$ leads to eternal accelerated expansion. Because
\mbox{$\lambda=2/\sqrt{6\alpha}$}, the corresponding bounds on $\alpha$ for 
transient accelerated expansion are

%The steepness of the exponential tail determines whether or not a quintessential inflation model is viable to successfully predict what is observed today. For the scalar field to account for the present dark energy component of the Universe and to avoid eternal acceleration we must impose $V_{\mathrm{atr}} \leq \rho_{\mathrm{B}}$ \cite{Kostas:1}. 
%Writing the potential in the form:
%
%\begin{equation}\label{eq:Quint_V_general}
%V(\phi \gg \phi_{\mathrm{end}}) \simeq V_{\mathrm{late}} \mathrm{exp}\Big(\frac{-\lambda\phi}{m_P}\Big)	\quad\,,
%\end{equation}
%where $\lambda$ contains the free parameters, leads us to the following constraints:
%
%\begin{equation}\label{eq:epsilon_constraint}
%\frac{V_{\mathrm{atr}}}{\rho_{\mathrm{B}}} < 1 \qquad \rightarrow \qquad \epsilon^2 > \frac{(1-\omega_B^2)}{4}	\quad\,.
%\end{equation}
%
%Remembering $\epsilon = \frac{m_P}{\sqrt{6}}\Big|\frac{V'}{V}\Big|$ and in this regime $V' = \frac{-\lambda}{m_P}V$: 
%
%\begin{equation}\label{key}
%\epsilon = \frac{\lambda}{\sqrt{6}}	\quad\,,
%\end{equation}
%and so the constraint in Eq.\eqref{eq:epsilon_constraint} becomes:
%
%\begin{equation}\label{eq:final_lambda_constraint}
%\lambda \geq \sqrt{\frac{3}{2}}	\quad\,,
%\end{equation}
%where we have used $\omega = 0$ for a matter dominated Universe. 
	
	\begin{equation}\label{eq:final_alpha_values}
0.03%\simeq\frac{1}{36} 
\lesssim \alpha \leq %\frac13\simeq 
0.33\quad,
%0.11 \leq \alpha \leq 0.44 
	\end{equation}
while for \mbox{$\alpha>\frac13$} we end up with eternal accelerated expansion.
However, as explained earlier, to avoid super-Planckian values for the 
non-canonically normalised $\phi$, we consider \mbox{$\alpha\lesssim\frac16$}. 
Thus, we see that, with \mbox{$\alpha\sim 0.1$} we attain transient accelerated
expansion with mildly sub-Planckian values of $\phi$.

Now, transient accelerated expansion has a clear merit over eternal accelerated
expansion, as in $\Lambda$CDM. This is the known future horizon problem of 
string theory. Eternal accelerated expansion results in a future event horizon.
As a result, future asymptotic states are not well defined because space is 
not causally connected. Thus, the S-matrix in string theory, which defines 
transition amplitudes, cannot be formulated \cite{horizons}. This may be just a 
problem of string theory and not a no-go theorem of nature. But still, it is an
incentive to avoid eternal acceleration if possible.

\subsection{The Parameter Space}

We now find the parameter space for $n$ and $\kappa$. To do this, we enforce 
the requirement that the density of quintessence $V_F$ must be comparable 
with the density of the Universe at present $\rho_0$ (coincidence requirement). 
We have
\begin{equation}
\frac{\rho_{\rm inf}}{\rho_0}\simeq\frac{V_{\rm inf}}{V_F}\simeq
\frac{e^{\lambda\varphi_F/m_P}}{2ne^{-2n}}\sim 10^{108}\quad,
%\quad\Rightarrow\quad
%\frac{e^{\lambda\varphi_F/m_P}}{2ne^{-2n}}\sim\left(\frac{}{}\right)
\end{equation}
where we used 
\mbox{$V_{\rm inf}=(1-e^{-2n})\,M^4\simeq M^4\sim(10^{15}\,{\rm GeV})^4$}
(we will find \mbox{$n\gg 1$}), \mbox{$\rho_0\sim(10^{-3}\,{\rm eV})^4$},
\mbox{$V_F=V_{\mathrm{Q}}\exp(-\lambda\varphi_F/m_P)$} and 
\mbox{$V_{\mathrm{Q}} = 2ne^{-2n}M^4$}. The above leads to

\begin{equation}\label{eq:lambda_with_phi}
%\lambda = \frac{\mathrm{ln}(2n) - 2n +110\mathrm{ln}(10)}{\varphi} = 
%\frac{\mathrm{ln}(2n) - 2n + 253}{\varphi}	\quad\,.
2n-\mathrm{ln}(2n)\simeq 108\ln10-\frac{2}{\sqrt{6\alpha}}\frac{\varphi_F}{m_P}
\quad,
\end{equation}
where we used $\lambda=2/\sqrt{6\alpha}$. Ignoring for the moment 
$\varphi_{\rm end}$, from Eq.~\eqref{eq:phi_f_3}, we have
\begin{equation}
\frac{\varphi_F}{m_P}\simeq\mbox{$\sqrt{\frac23}$}
\left(1-\mbox{$\frac32$}\ln\Omega_\gamma^{\rm end}\right)\quad,
\label{phiF}
\end{equation}
where, with gravitational reheating, we have 
\mbox{$\Omega_\gamma^{\rm end}=(\Omega_\gamma^{\rm end})_{\rm gr}$} given by 
Eq.~\eqref{greff}, which suggests
\mbox{$(\Omega_\gamma^{\rm end})_{\rm gr}\sim 10^{-3}(M/m_P)^4\sim 10^{-15}$},
where we used that \mbox{$M\sim\sqrt{H_{\rm inf}m_P}\sim 10^{15}\,$GeV}
(cf. Table~\ref{table:Nforalpha}).
Inserting this number into the above we find 
\mbox{$\varphi_F\simeq 43\,m_P$}. We put this into 
Eq.~\eqref{eq:lambda_with_phi} and find the allowed values of $n$.
For the range in Eq.~\eqref{eq:final_alpha_values},
%$\frac{1}{36}<\alpha\leq\frac16$, 
which corresponds to transient accelerated expansion with mildly
sub-Planckian $\phi$, we obtain
\begin{equation}\label{nrange}
25\lesssim n\leq 92
\quad.
\end{equation}
The corresponding values of $\varphi_{\rm end}$ can be obtained from 
Eq.~\eqref{eq:phi_e}. We find \mbox{$-3.20\leq\varphi_{\rm end}/m_P\lesssim-0.63$}
for the range of $\alpha$ considered. This substantiates our assumption to
ignore $\varphi_{\rm end}$ as \mbox{$|\varphi_{\rm end}|\ll\varphi_F$}.
In view of the above range, we also obtain 
\mbox{$V_0^{1/4}=e^{-n/4}M=10^{5-12}\,$GeV}, which is a quite large range of 
reasonable intermediate energy scales.

Using the values in Eq.~\eqref{nrange}, since 
\mbox{$\kappa\equiv n/\sqrt{6\alpha}$}, we readily obtain
\begin{equation}\label{krange}
59 \lesssim\kappa\leq 65\quad.
\end{equation}
Thus, because \mbox{$\kappa\approx 60$}, the non-canonical $\phi$ in the 
exponent of the scalar potential in Eq.~\eqref{eq:lagrangian1} is suppressed by 
the mass-scale \mbox{$m_P/\kappa\simeq 4\times 10^{16}\,{\rm GeV}\sim M$}.
We also find that the scale of the cosmological constant is
\mbox{$\Lambda^{1/4}=e^{-n/2}M\sim 10^{-5}-10^{10}\,$GeV}. In particular, for
\mbox{$\sqrt{6\alpha}\lesssim 1$} ($\phi\lesssim m_P$) we have 
\mbox{$n=\kappa\sqrt{6\alpha}\lesssim 60$}, so that 
\mbox{$\Lambda^{1/4}\gtrsim 10^2\,$GeV}, which is comparable with the electroweak energy scale.

As both the tensor-to-scalar ratio and the spectral index are independent of 
$n$, we simply use the constraints on $\alpha$ in 
Eq.~\eqref{eq:final_alpha_values} 
to update our results. For the spectral index we find a firm prediction
\begin{equation}\label{final_ns_values}
%0.96856\lesssim n_s\leq 0.96862
n_s=0.9686
\quad,
\end{equation}
which is very close to the value \mbox{$n_s=0.9685$} that we obtained for 
negligible $\alpha$ in Sec.~\ref{nsandr}. For the running of the spectral index,
we find the value \mbox{$n_s'=-5.09\times 10^{-4}$} which is virtually 
indistinguishable from the result obtaining in Sec.~\ref{nsandr} with 
negligible $\alpha$.

\begin{figure} [h]
	\centering
	\includegraphics[width=\linewidth]{%second_graph.png
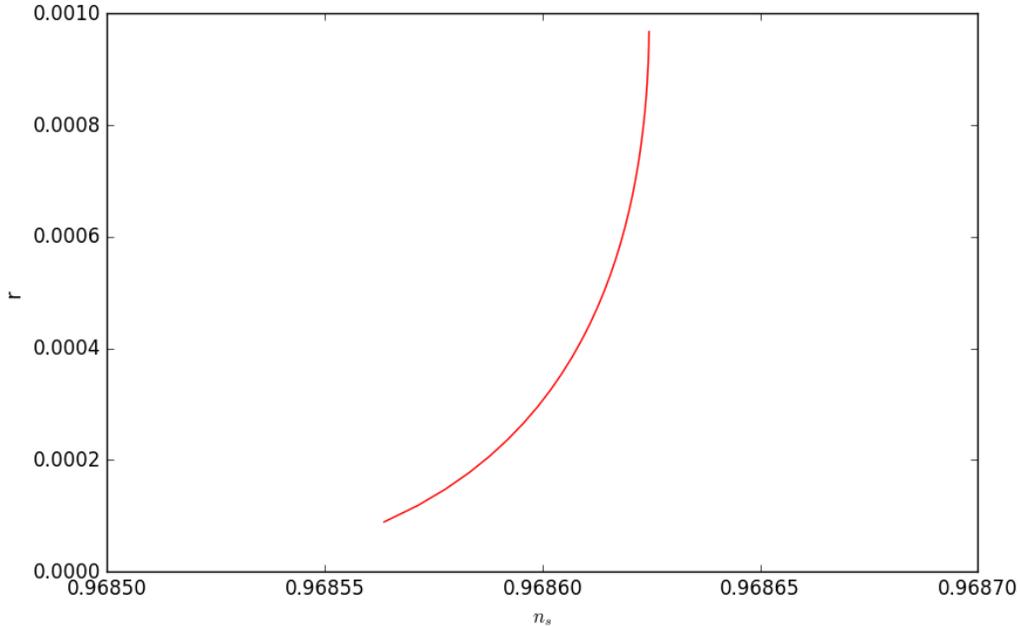} 
	\caption{The tensor-to-scalar ratio, $r$, versus the spectral index, 
$n_s$ for $\alpha = 0.03(0.33)$ left to right. All of these results are 
well within the Planck 1-$\sigma$ range. Values of $r\sim 10^{-3}$ are potentially observable in the near future.}  
	\label{fig:n_s_and_r2} 
\end{figure}

For $N_*=63.49$, Eq.~\eqref{eq:r} gives \mbox{$r\simeq\frac{\alpha}{340}$}. In 
view of Eq.~\eqref{eq:final_alpha_values} we find the following range of values 
for the tensor-to-scalar ratio:

\begin{equation}\label{final_r_values}
8.9\times 10^{-5}\lesssim r\leq 9.7\times 10^{-4}\quad.
\end{equation}
which can be potentially observable in the near future.
These results are plotted in Fig.~\ref{fig:n_s_and_r2}.

Since Ref.~\cite{trans2} is somewhat dated, we intend to revise the bound 
in Eq.~\eqref{eq:final_alpha_values} in a future publication, in view of the 
latest CMB constraints on the equation of state of dark energy. We expect that
the parameter space will be somewhat reduced. The predictions of our model, 
however, are rather robust. For example, this is evident from 
Fig.~\ref{fig:n_s_and_r2} regarding the value of the spectral index. The same 
is true for the model parameters, since \mbox{$\kappa\approx 60$} and $V_0$ and 
$\Lambda$ assume reasonable intermediate density scales in the parameter space
in Eq.~\eqref{eq:final_alpha_values}. 
%Still, the detailed study of the behaviour
%of the system today, when the quintessence field unfreezes and approaches its 
%late-time attractor, will provide estimates of the value of the barotropic 
%parameter of quintessence as a function of redshift and the duration of the 
%transient accelerated expansion. %, which are beyond the present paper.

%\clearpage

\section{Additional Considerations}

\subsection{The Fifth Force Problem and Radiative Corrections%
}\leavevmode

Our scalar field may be coupled to the standard model fields.
As is typical for a quintessence field, the mass of our field at present 
is of the order of the Hubble constant \mbox{$H_0=1.43\times 10^{-33}\,$eV}. This
means that its Compton wavelength is of the order of the size of the horizon and
the quintessence field can result in a long-range interaction (fifth force),
which is in danger of violating the equivalence principle. To quantify this,
any interaction terms in the Lagrangian density must be taken into account. Such
terms are of the form \cite{carroll}
\begin{equation}\label{inter}
\beta_i\,
%\Big(
\frac{\phi}{m_P}
%\Big)
\,\mathscr{L}_i
\end{equation}
where $i$ runs over the various different interactions, so that the 
dimensionless couplings $\beta_i$ may be different for each gauge-invariant 
dimension-four
operator $\mathscr{L}_i$ of the Lagrangian. For example, for the electromagnetic
field, the Lagrangian density can be of the form
\begin{equation}
\mathscr{L}_{\rm em}=-\frac14 e^{-4\beta_{\rm em}(\phi/m_P)}F_{\mu\nu}F^{\mu\nu}\simeq
-\frac14 F_{\mu\nu}F^{\mu\nu}+\beta_{\rm em}\frac{\phi}{m_P}F_{\mu\nu}F^{\mu\nu}\quad,
\end{equation}
where $F_{\mu\nu}$ is the field strength tensor \cite{eminter}. The above 
interaction can be responsible for variation of the fine-structure constant 
\cite{varyingA}.

Therefore, if the field is sub-Planckian, these interactions are suppressed. As 
we have seen, kination sends the canonical scalar field $\varphi$ to strongly 
super-Planckian values, which means that substantial fine-tuning of the 
$\beta_i$s is needed to avoid the so-called fifth force problem of traditional 
quintessence \cite{5th}. This is because, the above gravitationally suppressed 
interactions can result in potentially observable violations of the equivalence 
principle.

In our model, however, it is the non-canonical field $\phi$ which is expected 
to feature in the interaction terms of the form in Eq.~\eqref{inter}. Because, 
in our model, $\phi$ avoids being super-Planckian as long as 
$\sqrt{6\alpha} \leq 1$, the fifth force problem can be avoided with only a
mild tuning of the $\beta_i$ coefficients. At late times, in our model, we find 
the couplings to be 
\mbox{$\beta_i\sqrt{6\alpha}\mathscr{L}_i\sim\beta_i\mathscr{L}_i$},
where we considered \mbox{$\alpha\sim 0.1$}, as suggested by 
Eq.~\eqref{eq:final_alpha_values}. Therefore, for mildly suppressed values of 
$\beta_i$ transient accelerated expansion is possible without sizeable 
violations of the equivalence principle.

A related issue is the lifting of the flatness of the quintessential tail
by radiative corrections. Indeed, as we have seen, kination propels the field 
to a distance more that $40\,m_P$ in field space. A perturbative potential
is not valid over such super-Planckian field displacements, as one expects
non-renormalisable corrections of the form $\sim\varphi^{2i+4}/m_P^{2i}$ (with 
\mbox{$i\geq 1$}) to become important. Of course, ours is a 
non-perturbative potential (possibly originating from gaugino condensation)
but still we expect that the flatness of the quintessential tail would be 
threatened by non-renormalisable terms. However, in the context of 
$\alpha$-attractors, it is the non-canonical field $\phi$ (and not $\varphi$) we
should be concerned about, because this is the fundamental degree of freedom 
that appears in our original Lagrangian in Eq.~\eqref{eq:lagrangian1}, while 
the canonically normalised $\varphi$, introduced in 
Eq.~\eqref{eq:canonicalVariables} is merely a mathematical tool to help 
study the model. As we saw, the variation of $\phi$ is kept mildly
sub-Planckian (\mbox{$\alpha\sim 0.1$}), which means that radiative corrections
are kept under control. %(with only a mild tuning of the $C_i$ coefficients).

\subsection{Overproduction of Gravitinos}
\label{grav}

Our model is rooted in supergravity, since scalar fields with non-canonical 
kinetic terms naturally arise in this framework. Therefore, one limitation 
which should be taken into account is the overproduction of gravitinos. 

The gravitino is the super-partner of the graviton and it is expected to have a 
mass of the order of TeV. This presents two options: Either the gravitino is 
stable (e.g. being the lightest supersymmetic particle) and its mass 
contributes to dark matter, in which case overproduction of gravitinos 
overcloses the Universe.
%This is a problem though because the observed density is not as high as the density of the gravitinos - hence the term 'overproduction of gravitinos'. 
Or, the gravitino is unstable and decays to other particles. Gravitinos can only
decay via gravitationally suppressed interactions, which progress very slowly, 
giving them a long life-time and meaning they exist past the time of BBN.
However, the channels of decay a gravitino can use, necessarily produce 
particles energetic enough to destroy the nuclei created during BBN. 
Hence, we need to avoid gravitino overproduction entirely.
 
The gravitinos can be produced by either a thermal or non-thermal mechanism. 
Because the inflaton does not oscillate around its VEV in our model, we 
are only concerned with the thermal production - their creation via scatterings 
in the thermal bath produced via gravitational reheating. The relative abundance
of produced gravitinos depends strongly on the reheating temperature 
$T_{\mathrm{reh}}$, i.e. the relevant temperature in the transition to
radiation domination. To avoid overproduction, $T_{\rm reh}$ is constrained 
to be below $10^{8}-10^{9}\,\mathrm{GeV}$ \cite{Constraints_on_T_reh} 
(but sometimes it can be much lower than that; of order $10^6\,$GeV or so 
\cite{kaz}.)

From Eq.~\eqref{Treh}, for gravitational reheating we have
\begin{equation}\label{eq:T_reh}
T_{\rm reh}=\frac{q^{3/4}}{36\pi^2}
\left(\frac{g_*^{\rm end}}{g_*^{\rm reh}}\right)^{1/4}
\sqrt{\frac{g_*^{\rm end}}{10}}\,\frac{V_\mathrm{end}}{m_P^3}\quad,
\end{equation}
%\begin{equation}\label{key}
%T_{\mathrm{reh}} = \frac{1}{72\pi^2}\sqrt{\frac{g_*}{30}}
%\frac{V_\mathrm{end}}{m_P^3}
%\end{equation}
where $V_{\mathrm{end}}\equiv V(\varphi_{\mathrm{end}})$ and we have taken
\mbox{$V_{\rm end}\simeq\frac32H_{\rm end}^2m_P^2$}. Using Eqs.\eqref{eq:Infl_pot} 
and \eqref{eq:phi_e} we find that \mbox{$V_{\rm end}=e^{-\sqrt{3\alpha}}M^4$}. Then,
taking %\mbox{$g_*^{\rm end}=106.75$} 
\mbox{$g_*^{\rm end}={\cal O}(100)$} 
and \mbox{$g_*^{\rm reh}=10.75$}, and 
considering also Eq.~\eqref{eq:final_alpha_values} and
Table~\ref{table:Nforalpha} %(for the values of $M$) 
we find \mbox{$T_{\rm reh}\sim 10^4\,$GeV}
(cf. also Sec.~\ref{reh}). Thus, $T_{\mathrm{reh}}$ is well below the upper-bound 
to avoid gravitino overproduction. Moreover, because 
$T_{\mathrm{reh}}\gg T_{\mathrm{BBN}}$, where $T_{\mathrm{BBN}}\simeq 0.5\,\mathrm{MeV}$
is the temperature at BBN, we find that latter is safely not affected.

\subsection{Overproduction of Gravitational Waves}

The non-decaying mode for a gravitational wave with superhorizon wavelength has 
a constant amplitude until it re-crosses the horizon and then it undergoes 
damped oscillations, decreasing as $1/a$. This is a general result valid for 
any equation of state, however because the damping is affected by the scale 
factor the energy density spectrum of the gravitational waves scales differently
in different epochs of Universe expansion. During the radiation era, the 
barotropic parameter of the Universe is $w=\frac13$ and the gravitational wave
spectrum is flat. However, during kination we have $w=1$ which produces a spike 
in the spectrum of gravitational waves at high frequencies. In order to ensure 
the generated gravitational waves do not destabilise BBN, an upper bound is
imposed on their density fraction~\cite{Maggiore:1999vm}: 

\begin{equation}\label{eq:GW_bound}
I \equiv h^2 \int_{k_{\mathrm{BBN}}}^{k_{\mathrm{end}}}\Omega_{\mathrm{GW}}(k)d\mathrm{ln}k \leq 1\times 10^{-5} 	\qquad \,,
\end{equation}
where $h = 0.678$.
%Three different behaviours are found for modes which entered the Hubble radius during the kination, radiation domination and matter domination regimes and 
We focus on the modes which re-enter the horizon during the kination, 
i.e. the spike in the spectrum, as this is the dominant contribution to 
$\Omega_{\mathrm{GW}}$. In this regime, we have~\cite{Gravitational Wave Spectrum}

%\begin{equation}\label{eq:grav_waves}
%\Omega_{\mathrm{GW}}(k) = \left\{
%								\begin{array}{ll}
%								\varepsilon\Omega_{\gamma}(k_0)h_{\mathrm{GW}}^2(\frac{k}{k_*})[\mathrm{ln}(\frac{k}{k_\mathrm{end}})]^2 \qquad	\qquad \quad k_*<k\leq k_{\mathrm{end}} \\
%								\\
%								\frac{\pi}{4}\varepsilon\Omega_{\gamma}(k_0) h_{\mathrm{GW}}^2[\mathrm{ln}(\frac{k_*}{k_{\mathrm{end}}})]^2	\qquad \qquad \qquad	k_{\mathrm{eq}} <k \leq k_*  \\
%								\\
%								\frac{\pi}{16}\varepsilon\Omega_{\gamma}(k_0)h_{\mathrm{GW}}^2(\frac{k_{\mathrm{eq}}}{k})^2 \mathrm{ln}(\frac{k_*}{k_{\mathrm{end}}})]^2	\qquad \qquad	k_0 <k \leq k_{\mathrm{eq}}
%								\end{array}
%								\right\}
%\end{equation}

\begin{equation}\label{eq:grav_waves}
\Omega_{\mathrm{GW}}(k) = 
\varepsilon\Omega_{\gamma}(k_0)h_{\mathrm{GW}}^2\Big(\frac{k}{k_{\mathrm{reh}}}\Big)
\Big[\mathrm{ln}\Big(\frac{k}{k_\mathrm{end}}\Big)\Big]^2 
\qquad{\rm for}\quad k_{\mathrm{reh}}<k\leq k_{\mathrm{end}}\quad,
\end{equation}
where 
%subscipts `end' and `reh' refer to the end of inflation and the end of kination respectively. 
$k$ is the mode's physical momentum and $\Omega_{\gamma}(k_0)$ is the present 
density fraction of radiation, on horizon scales. We also have
\mbox{$h_{\mathrm{GW}}^2 = \frac{1}{8\pi}\big(\frac{H_{\mathrm{end}}}{m_P}\big)^2$} 
and \mbox{$\varepsilon=2R_i\big(\frac{g_{\mathrm{dec}}}{g_{\rm th}}\big)^{1/3}$},
where \mbox{$R_i = \frac{81}{32 \pi^3} $} is the contribution of each massless
scalar degree of freedom to the energy density of the amplified 
fluctuations~\cite{Gravitational Wave Spectrum}. Using Eq.\eqref{eq:grav_waves} 
in \eqref{eq:GW_bound} we calculate

\begin{equation}%\label{key}
I=\varepsilon h^2 \Omega_{\gamma}(k_0) h_{\mathrm{GW}}^2 
\Big\{ 2\Big(\frac{k_{\mathrm{end}}}{k_{\mathrm{reh}}}\Big) - 
\Big[\mathrm{ln}\Big(\frac{k_{\mathrm{end}}}{k_{\mathrm{reh}}}\Big)+1\Big]^2\Big\}	
\qquad.
\end{equation}
Because $k_{\mathrm{end}} \gg k_{\mathrm{reh}}$ the first term in the brackets 
dominates and we simply obtain

\begin{equation}%\label{key}
I\simeq 2\varepsilon h^2\Omega_{\gamma}(k_0)h_{\mathrm{GW}}^2
\Big(\frac{k_{\mathrm{end}}}{k_{\mathrm{reh}}}\Big)	\qquad.
\end{equation} 
Since $k=aH$ and during kination $a \propto \rho^{-1/6}$ we have:
\begin{equation}%\label{key}
I = 2 \varepsilon h^2 \Omega_{\gamma}(k_0) h_{\mathrm{GW}}^2 
\Big(\frac{\rho_{\mathrm{reh}}}{\rho_{\mathrm{end}}}\Big)^{1/6} 
\Big(\frac{H_{\mathrm{end}}}{H_{\mathrm{reh}}}\Big)	\qquad,
\end{equation} 
%With the relations $H_i^2 = \frac{2\rho_i}{3m_P^2}$ where $i =$ `end' or `reh' and equating 
Employing that \mbox{$\frac12\rho_{\mathrm{reh}}\approx\rho_\gamma^{\mathrm{reh}}
=\frac{\pi^2}{30}g_{*}^{\mathrm{reh}}T_{\mathrm{reh}}^4$} and 
\mbox{$\rho_{\mathrm{end}}\approx 2V_{\mathrm{end}}$} we obtain

\begin{equation}%\label{key}
I = \frac{2\varepsilon h^2\Omega_{\gamma}(k_0)}{\pi^{2/3}}
\left(\frac{30}{g_{*}^{\mathrm{reh}}}\right)^{1/3}
\frac{h_{\mathrm{GW}}^2V_{\mathrm{end}}^{1/3}}%
{T_{\mathrm{reh}}^{4/3}}		\qquad \,.
\end{equation}
Inserting Eq.~\eqref{eq:T_reh} we find

\begin{equation}%\label{key}
I = %45\pi\,2^{8/3} 
\frac{360\pi}{2^{1/3}}
\frac{\varepsilon h^2 \Omega_{\gamma}(k_0) } 
{q\, g_{*}^{\mathrm{end}}}		\qquad \,,
\end{equation}
where we have substituted in the value of $h_{\mathrm{GW}}^2$. Using 
$\Omega_{\gamma}(k_0) = 2.6\times 10^{-5}h^{-2}$ and introducing 
the values for the remaining constants $\varepsilon$ and 
\mbox{$g_*^{\rm end}=106.75\,{\cal C}$}, the bound of $I \leq 1\times 10^{-5}$ in 
Eq.~\eqref{eq:GW_bound} gives

\begin{equation}%\label{key}
q \geq 3.54/{\cal C}		
\qquad \,,
\end{equation}
where ${\cal C}=1$ for the standard model but ${\cal C}\gtrsim 2$ for 
supersymmetric theories.
Hence, a reheating efficiency  $q={\cal O}(1)$ satisfies the bound in 
Eq.~\eqref{eq:GW_bound}, which means that BBN remains undisturbed from 
graviational reheating, as we have assumed.
%and the results for $T_{\mathrm{reh}}$ in the previous section are valid.

\section{Discussion and Conclusions}

We have presented and analysed a new model of quintessential inflation within 
the context of $\alpha$-attractors, motivated by supergravity theories. We have
assumed a simple exponential potential for the non-canonical inflaton field, 
which may originate from gaugino condensation. We also, considered a vanishing 
vacuum density, due to some unknown symmetry, as was standard practice before 
the observation of dark energy. However, this does not imply a zero cosmological
constant, because the value of our scalar field cannot go to infinity (where 
its exponential potential density would tend to zero) due to a bound imposed
by a pole in the non-canonical kinetic term. The energy scale found for the 
cosmological constant ($\gtrsim 10^2\,$GeV) is comparable to the electroweak
energy scale, in stark contrast with the required value in $\Lambda$CDM 
($\sim 10^{-3}\,$eV).

Our model performs very well both as an inflation model and as quintessence.
The inflationary observables are near the sweet spot of the latest CMB 
observations, with the spectral index found as \mbox{$n_s\simeq 0.9686$} or so
and the tensor-to-scalar ratio as \mbox{$r\sim 10^{-4}-10^{-3}$} that is 
potentially observable in the near future. These values lie deep inside the 
1-$\sigma$ contour of the Planck and BICEP2 results. This is not surprising for 
a plateau inflationary model. For quintessence, we have shown that the model 
leads to transient accelerated expansion at present for rather natural values of
the model parameters. Indeed, we found that \mbox{$\alpha\sim 0.1$} and 
\mbox{$\kappa\sim m_P/M$} in Eq.~\eqref{eq:lagrangian1}, with $M$ being the 
inflation energy scale, which is close to the energy of grand unification.
We also found a large but reasonable range of intermediate density scales for 
$V_0$ (ranging as \mbox{$V_0^{1/4}\sim 10^{5-12}\,$GeV}).

Transient accelerated expansion improves over the usual eternal accelerated 
expansion of $\Lambda$CDM in that it does not lead to a future horizon problem,
which otherwise undermines the formulation of the S-matrix in string theories.
It also results in an ultimate future for our Universe different from the
$\Lambda$CDM scenario. In $\Lambda$CDM, eternal accelerated expansion results
to all unbound or loosely bound systems (like galactic clusters) eventually 
dissolving and pulled beyond the constant horizon, while our galaxy, merged with
all other objects in the local group, remains the only object surviving 
within our observable Universe, which will be filled with Hawking radiation at 
temperature \mbox{$T\sim H_0\sim 10^{-33}\,$eV}. In contrast, with transient 
accelerated expansion, the horizon will continue to expand with the speed of 
light, while ever more objects will enter the observable Universe. All mass 
concentrations most probably will eventually become black holes that will 
evaporate in diminishing radiation with temperature that will asymptote to 
zero.

Ours is a non-oscillatory inflation model, so that the inflaton field does not 
decay at the end of inflation but survives until today to become quintessence. 
Therefore, we need to reheat the Universe by means other than the decay of the 
inflaton field.
In an effort to keep our model minimal, we have not utilised the help of other
degrees of freedom, which would play a crucial role in reheating the Universe. 
Instead, we considered gravitational reheating, which is based on particle 
production of all light (and not-conformally invariant) fields during inflation.
Such production always occurs but it is negligible in the usual oscillatory 
models of inflation. Hence, gravitational reheating is a neat mechanism to 
reheat the Universe, since it is unavoidable. We find the reheating temperature
\mbox{$T_{\rm reh}\sim 10^4\,$GeV}, which means that reheating occurs safely  
before big bang nucleosynthesis. Also, our $T_{\rm reh}$ satisfies even the most 
stringent constraints due to gravitino overproduction.

Between inflation and reheating, there is a period of kination, when the 
density of the Universe is dominated by the kinetic density of our scalar field.
We have studied kination in a model independent way. We showed that, soon after
kination ends, the scalar field freezes at a constant value, where it remains 
dormant until late times, when it can become quintessence. With gravitational 
reheating, we found that the displacement, during kination, of the canonical 
field is strongly super-Planckian $\sim 45\,m_P$. This would have meant that 
the flatness of our quintessential plateau could be lifted by radiative 
corrections, while the long-range fifth force mediated by quintessence could 
lead to violations of the equivalence principle. In the context of 
$\alpha$-attractors, though, since the non-canonical field may avoid being
super-Planckian,
the above dangers can be avoided with only a mild tuning of the gravitationally 
suppressed couplings between quintessence and the standard model fields.
The period of kination results in a spike in the spectrum of gravitational 
waves generated during inflation. We have calculated this spectrum and found 
that the energy of the gravitational waves is not threatening big bang
nucleosynthesis.
%while the frequency and amplitude of the spike are such that they may be 
%observable by LIGO.

In summary, we have studied a new quintessential inflation model, in the 
context of $\alpha$-attractors. We have found excellent agreement of the 
inflationary predictions with CMB observations. We have shown that successful
quintessence is achieved with natural values of the parameters, including a 
cosmological constant near the electroweak energy scale. Our model gives rise to
transient accelerated expansion, dispensing with the future horizon problem of 
$\Lambda$CDM. Our setup is purposefully minimal, without many degrees of 
freedom, mass scales and couplings. We considered gravitational reheating, which
does not overproduce gravitinos nor does it affect nucleosynthesis. Finally, a 
period of kination in our model produces a spike in the spectrum of 
gravitational waves generated during inflation, which does not disturb big bang
nucleosynthesis. %but it is potentially observable by LIGO.

\paragraph{Acknowledgements}\leavevmode\\
%
%\noindent
CO is supported by the FST of Lancaster University. KD is supported (in part) by
the Lancaster-Manchester-Sheffield Consortium for Fundamental Physics under 
STFC grant: ST/L000520/1

%\clearpage

\end{document}